\documentclass[letter]{jpsj2} 
\title{EuFe$_2$As$_2$ under High Pressure: an Antiferromagnetic Bulk Superconductor}

\author{Taichi \textsc{Terashima}$^{1, 4}$, Motoi \textsc{Kimata}$^{1}$, Hidetaka \textsc{Satsukawa}$^{1}$, Atsushi \textsc{Harada}$^{1}$, Kaori \textsc{Hazama}$^{1}$, Shinya \textsc{Uji}$^{1, 4}$, Hiroyuki S. \textsc{Suzuki}$^{2}$, Takehiko \textsc{Matsumoto}$^{2}$, and Keizo \textsc{Murata}$^{3}$}

\inst{$^{1}$National Institute for Materials Science, Tsukuba, Ibaraki 305-0003\\
$^{2}$National Institute for Materials Science, Tsukuba, Ibaraki 305-0047\\
$^{3}$Division of Molecular Materials Science, Graduate School of Science, Osaka City University, Osaka 558-8585\\
$^{4}$JST, Transformative Research-Project on Iron Pnictides (TRIP), Chiyoda, Tokyo 102-0075}

\abst{We report the ac magnetic susceptibility $\chi_{ac}$ and resistivity $\rho$ measurements of EuFe$_2$As$_2$ under high pressure $P$.  By observing nearly 100\% superconducting shielding and zero resistivity at $P$ = 28 kbar, we establish that $P$-induced superconductivity occurs at $T_c \sim$~30 K in EuFe$_2$As$_2$.  $\rho$ shows an anomalous nearly linear temperature dependence from room temperature down to $T_c$ at the same $P$.  $\chi_{ac}$ indicates that an antiferromagnetic order of Eu$^{2+}$ moments with $T_N \sim$~20 K persists in the superconducting phase.  The temperature dependence of the upper critical field is also determined.}

\kword{iron pnictides, pressure-induced superconductivity, susceptibility, upper critical field}

\begin{document}
\maketitle

\def\degree{\kern-.2em\r{}\kern-.3em}

The discovery of superconductivity (SC) at a transition temperature $T_c$ = 26 K in LaFeAsO$_{1-x}$F$_x$by Kamihara \textit{et al}.\cite{Kamihara08JACS} has triggered extensive studies of SC in layered iron pnictides and related compounds.  Rotter \textit{et al}. found that BaFe$_2$As$_2$ with a simpler structure can be made superconducting by doping: $T_c$ = 38 K in (Ba$_{1-x}$K$_x$)Fe$_2$As$_2$ with $x$ = 0.4.\cite{Rotter08PRL}  Perhaps more importantly, it is reported that 122 compounds of the form $A$Fe$_2$As$_2$ ($A$ = Ca, Sr, Ba, and Eu) can be tuned to SC by the application of high pressure $P$.\cite{Torikachvili08PRL, Park08JPCM, Alireza09JPCM, Fukazawa08JPSJ, Kotegawa09JPSJ, Igawa09JPSJ,  Lee08condmat, Miclea08condmat}  $P$ tuning can provide opportunities to determine the nature of the iron-pnictide high-temperature SC without being adversely affected by disorder due to doping.  However, most of these reports are based only on resistivity $\rho$ measurements and hence cannot establish the bulk nature of $P$-induced SC.\cite{Matsubayashi09condmat}  Even when magnetic measurements are reported, results are not conclusive:  In ref.~\citen{Alireza09JPCM}, magnetic measurements were performed on SrFe$_2$As$_2$ and BaFe$_2$As$_2$, but the observed volume fraction was expressed in arbitrary units.  In ref.~\citen{Lee08condmat}, the volume fraction of the $P$-induced superconducting phase of CaFe$_2$As$_2$ was estimated to be at least 50\%, while in ref.~\citen{Yu08condmat} CaFe$_2$As$_2$ was reported not to exhibit SC under hydrostatic $P$ produced by the use of helium as a pressure-transmitting medium.

EuFe$_2$As$_2$ exhibits two phase transitions, at $T_o\sim$ 190 K and $T_N\sim$ 19 K, at ambient $P$.\cite{Raffius93JPCS, Tegel08JPCM, Ren08PRB, Jeevan08PRB}  The transition at $T_o$ is a combined structural and magnetic transition, similar to those in the other 122 compounds: the crystal structure changes from tetragonal to orthorhombic and the Fe$^{2+}$ moments order antiferromagnetically.  The transition at $T_N$ is due to the antiferromagnetic (AFM) ordering of the Eu$^{2+}$ moments.  The AFM coupling of the Eu$^{2+}$ moments is rather weak: the field-induced paramagnetic state with a saturated moment of $\sim$7 $\mu_B$/Eu is easily reached by the application of $\sim$1 or 2 T in the $ab$-plane or along the $c$-axis, respectively.\cite{Jiang09NJP}  A temperature ($T$)-$P$ phase diagram has been determined from $\rho$ measurements:\cite{Miclea08condmat} while $T_o$ decreases with $P$ and is not detected above $P$ = 23 kbar, $T_N$ is nearly $P$-independent up to 26 kbar (the highest $P$ in ref.~\citen{Miclea08condmat}).  The authors of ref.~\citen{Miclea08condmat} state that $P$-induced SC at $T_c\sim$30 K occurs above 20 kbar.  However, their $\rho$ data (at $P$ = 21.6 kbar) shows only a partial drop and approximately half of the normal-state $\rho$ appears to remain as $T\to0$.  Obviously, further experimental confirmation is necessary.  

In this letter, we report measurements of the ac magnetic susceptibility $\chi_{ac}$ and $\rho$ of EuFe$_2$As$_2$ single crystals under high $P$.  By observing a nearly 100\% shielding volume fraction and a sharp resistive transition to the zero-resistivity state at $P$ = 28 and 29 kbar, we establish that EuFe$_2$As$_2$ is a bulk superconductor at these values of $P$.  We also show evidence that the AFM order of the Eu$^{2+}$ moments persists in the superconducting phase. 

A single-crystal ingot of EuFe$_2$As$_2$ was grown by the Bridgman method from a stoichiometric mixture of its constituent elements.  A $^3$He/$^4$He dilution refrigerator or a $^3$He refrigerator and a superconducting magnet were used for measurements.  For the measurements of $\chi_{ac}$ ($\chi^{\prime}-i\chi^{\prime\prime}$), a piece with an $ab$-plane area of 1.15 x 1.15 mm$^2$ and a $c$-axis thickness of 0.5 mm was cut.  Note that the use of a thick sample enabled us to reliably estimate the volume fraction for the external field $B_{appl}$ in the $c$-direction.  In the case of a thin sample, a large demagnetization factor for $B_{appl}\parallel c$ makes the estimation of the volume fraction very difficult.  For measurements along the $c$-axis (in the $ab$-plane), the sample was placed in a clamped piston-cylinder pressure cell with the $c$-axis (the $ab$-plane) parallel to the cylinder axis.  An ac modulation field ($f$ = 67.1 Hz and $B_{ac}$ $\sim$ 0.04~mT) and external magnetic field $B_{appl}$ were applied along the cylinder axis.  In order to estimate the size of the signal corresponding to 100\% shielding, a piece of Pb with nearly the same dimensions as the sample was measured with the same pick-up coil for the two orientations (namely, the $c$-axis and $ab$-plane orientations).  The accuracy of these estimations was estimated to be about $\pm$10\%.  For $\rho$ measurements, a thin sample with dimensions of $\sim$1 x 0.2 x 0.03 mm$^3$ was exfoliated, where 0.03 mm is along the $c$-axis.  After four gold wires were attached to a (001) surface with conducting silver paste, the sample was placed in a clamped piston-cylinder pressure cell with the longest axis, which is the electrical current direction, parallel to the cylinder axis.  A standard four-contact method was used with a low-frequency ac current ($I$ = 0.1 mA, $f$ = 17 Hz), and the field $B_{appl}$ was applied parallel to the current.  For both $\chi_{ac}$ and $\rho$ measurements, Daphne7474 (Idemitsu Kosan Co., Ltd., Tokyo) was used as a pressure-transmitting medium.\cite{Daphne7474}  This oil does not solidify up to 37 kbar at room temperature (RT)\cite{Murata08RSI} and hence ensures hydrostatic-pressure generation in the present measurements (highest $P$ $\sim$30 kbar).  Furthermore, the pressure cells were cooled slowly ($\leqslant$1.5 K/min) from RT down to $\sim$20 K to prevent the possible development of nonhydrostaticity.  Note that we always refer to the applied field $B_{appl}$, which may be different from the internal field by $\sim$1 T, because of the large saturation moment of Eu$^{2+}$.

\begin{figure}[tb]
\begin{center}
\includegraphics[width=6.5cm]{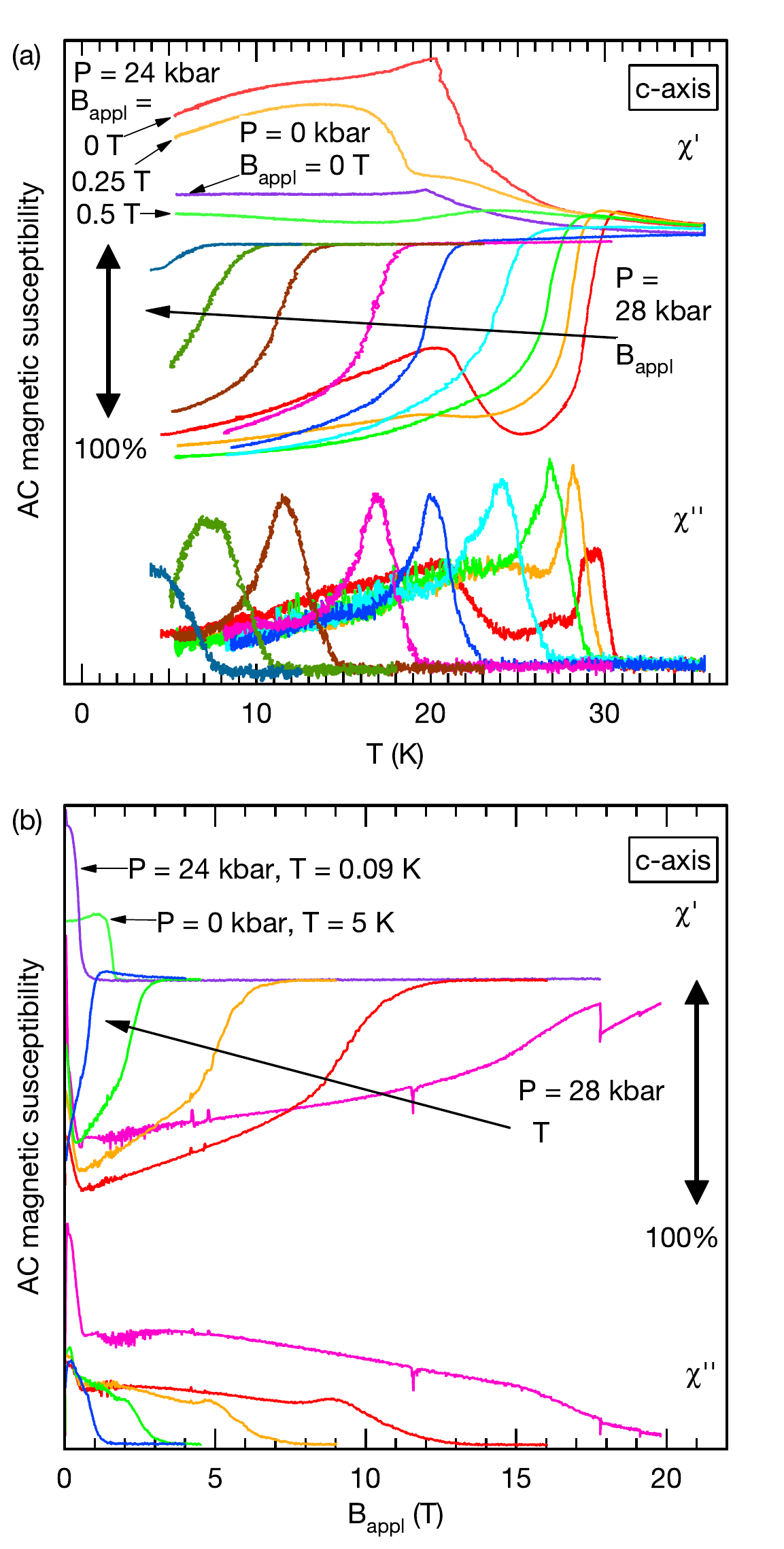}
\end{center}
\caption{(Color online) AC magnetic susceptibility $\chi^{\prime}-i\chi^{\prime\prime}$ along the $c$-axis of EuFe$_2$As$_2$ for $P$ = 0, 24, and 28 kbar (a) as a function of $T$ at a constant $B_{appl}$ and (b) as a function of $B_{appl}$ at a constant $T$.   $B_{appl}$ was applied parallel to the $c$-axis.  For $\chi^{\prime\prime}$ in (a), only data for $P$ = 28 kbar are shown.  The vertical lines with arrows at both ends indicate the estimated change in $\chi^{\prime}$ corresponding to a 100\% shielding volume fraction.  In (a), $B_{appl}$ = 0, 0.25, 0.5, 1, 2, 4, 8, 12, and 16 T for $P$ = 28 kbar.  In (b), $T$ = 0.02, 10, 15, 20, and 25 K for $P$ = 28 kbar.  During the measurement of the $T$ = 0.02 K curve at $P$ = 28 kbar in (b), the field was kept at $B_{appl}$ = 17.8 T for about 100 minutes for a technical reason, which caused a drop in $\chi^{\prime}$ for unknown reasons.}
\label{chi_c}
\end{figure}

\begin{figure}[tb]
\begin{center}
\includegraphics[width=6.5cm]{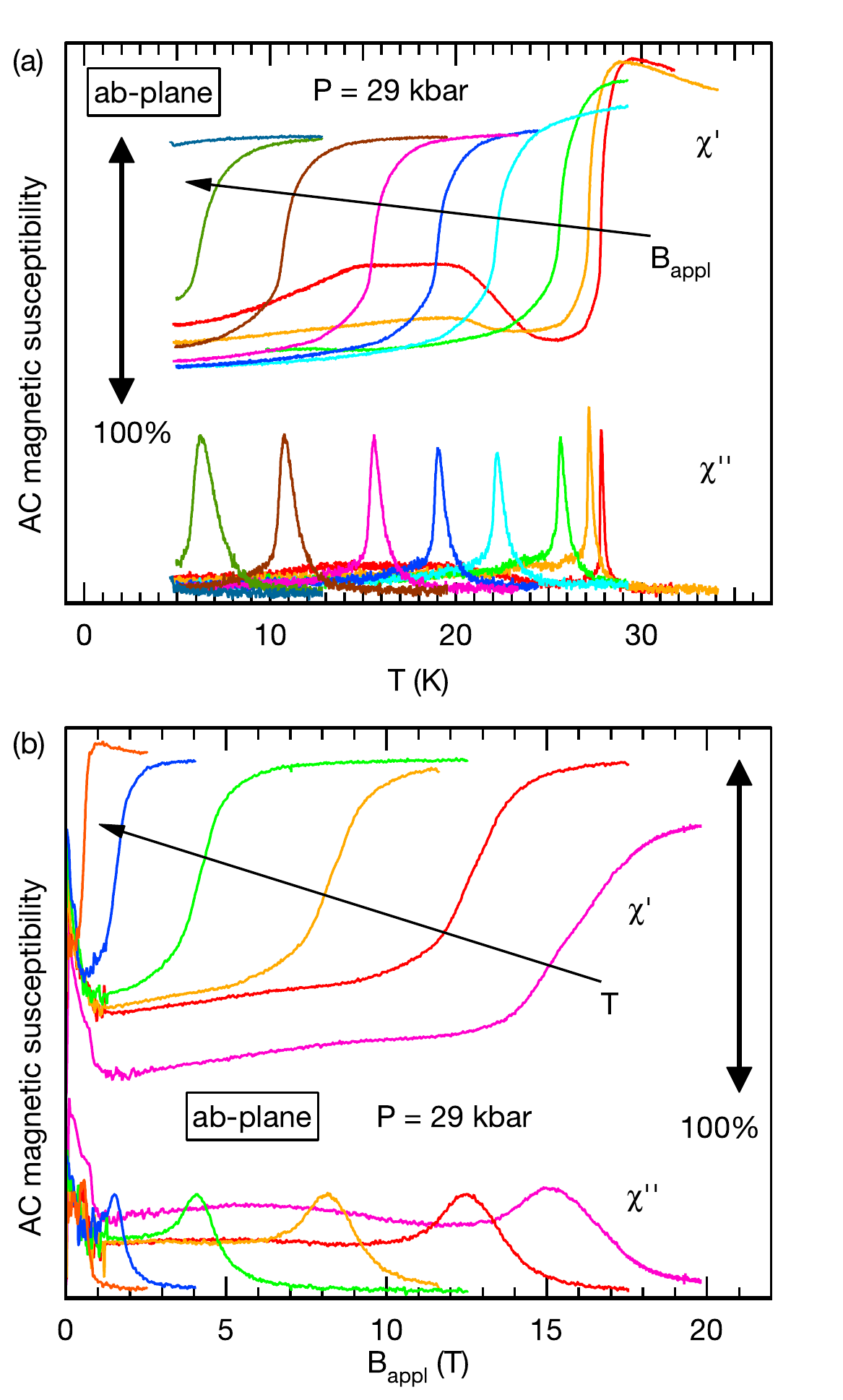}
\end{center}
\caption{(Color online) AC magnetic susceptibility $\chi^{\prime}-i\chi^{\prime\prime}$ in the $ab$-plane of EuFe$_2$As$_2$ for $P$ = 29 kbar as a function of (a) $T$ and (b) $B_{appl}$.   $B_{appl}$ was applied in the $ab$-plane.  The vertical lines with arrows at both ends indicate the estimated change in $\chi^{\prime}$ corresponding to a 100\% shielding volume fraction.  In (a), $B_{appl}$ = 0, 0.25, 0.5, 1, 2, 4, 8, 12, and 16 T.  In (b), $T$ = 0.3, 5, 10, 15, 20, and 25 K.}
\label{chi_ab}
\end{figure}

\begin{figure}[tb]
\begin{center}
\includegraphics[width=6.5cm]{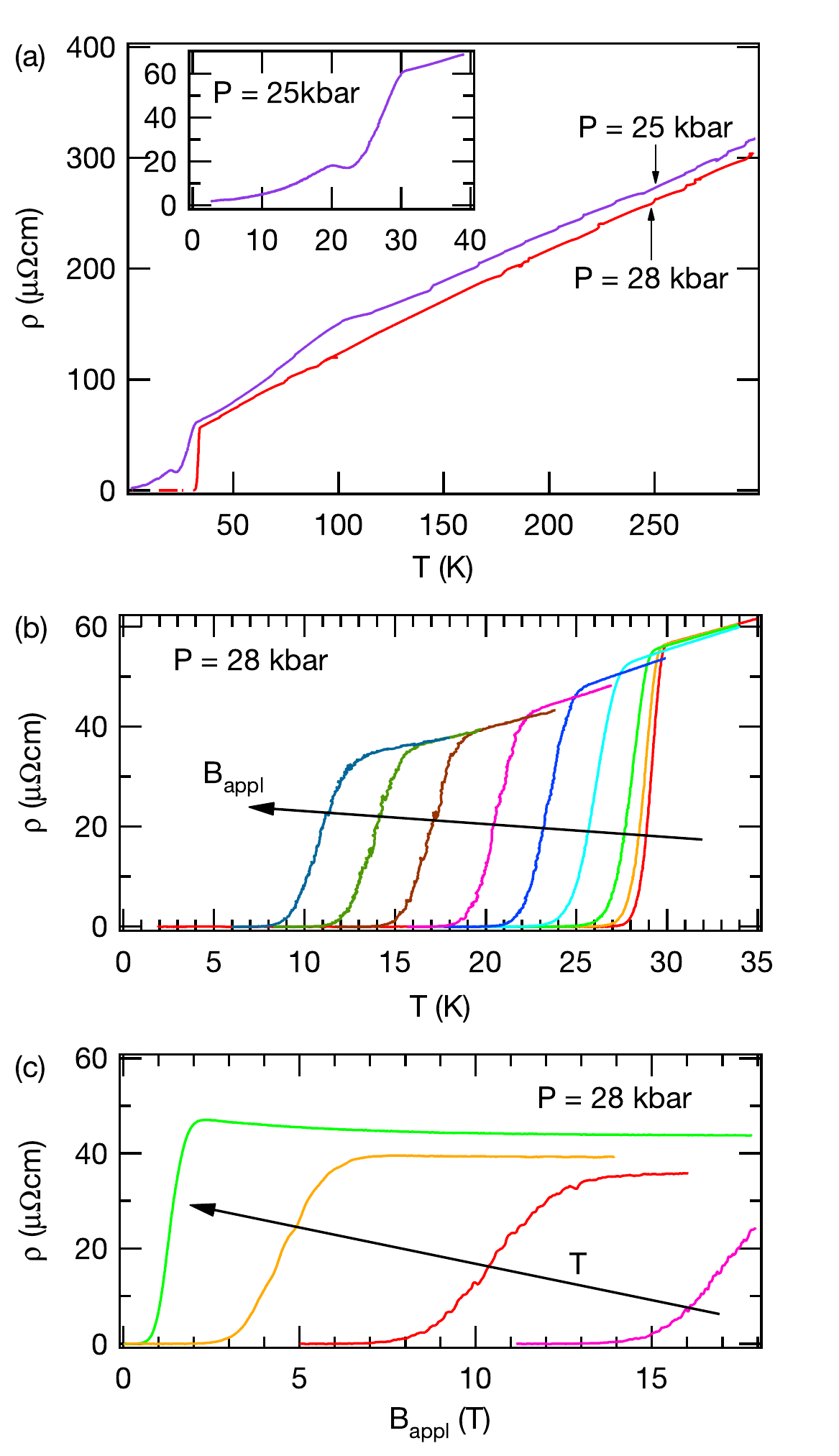}
\end{center}
\caption{(Color online) In-plane $\rho$ of EuFe$_2$As$_2$.  (a) $T$ dependence from RT at $P$ = 25 and 28 kbar.  The inset shows $\rho$ at $P$ = 25 kbar below $T$ = 40 K.  (b) $T$ dependence of $\rho$ at $P$ = 28 kbar near the superconducting transition.  $B_{appl}$ was applied parallel to the electrical current in the $ab$-plane: $B_{appl}$ = 0, 0.25, 0.5, 1, 2, 4, 8, 12, and 16 T.  (c) $B_{appl}$ dependence of $\rho$ at $P$ = 28 kbar for $T$ = 10, 15, 20, and 25 K.}
\label{rho_curves}
\end{figure}

\begin{figure}[tb]
\begin{center}
\includegraphics[width=7cm]{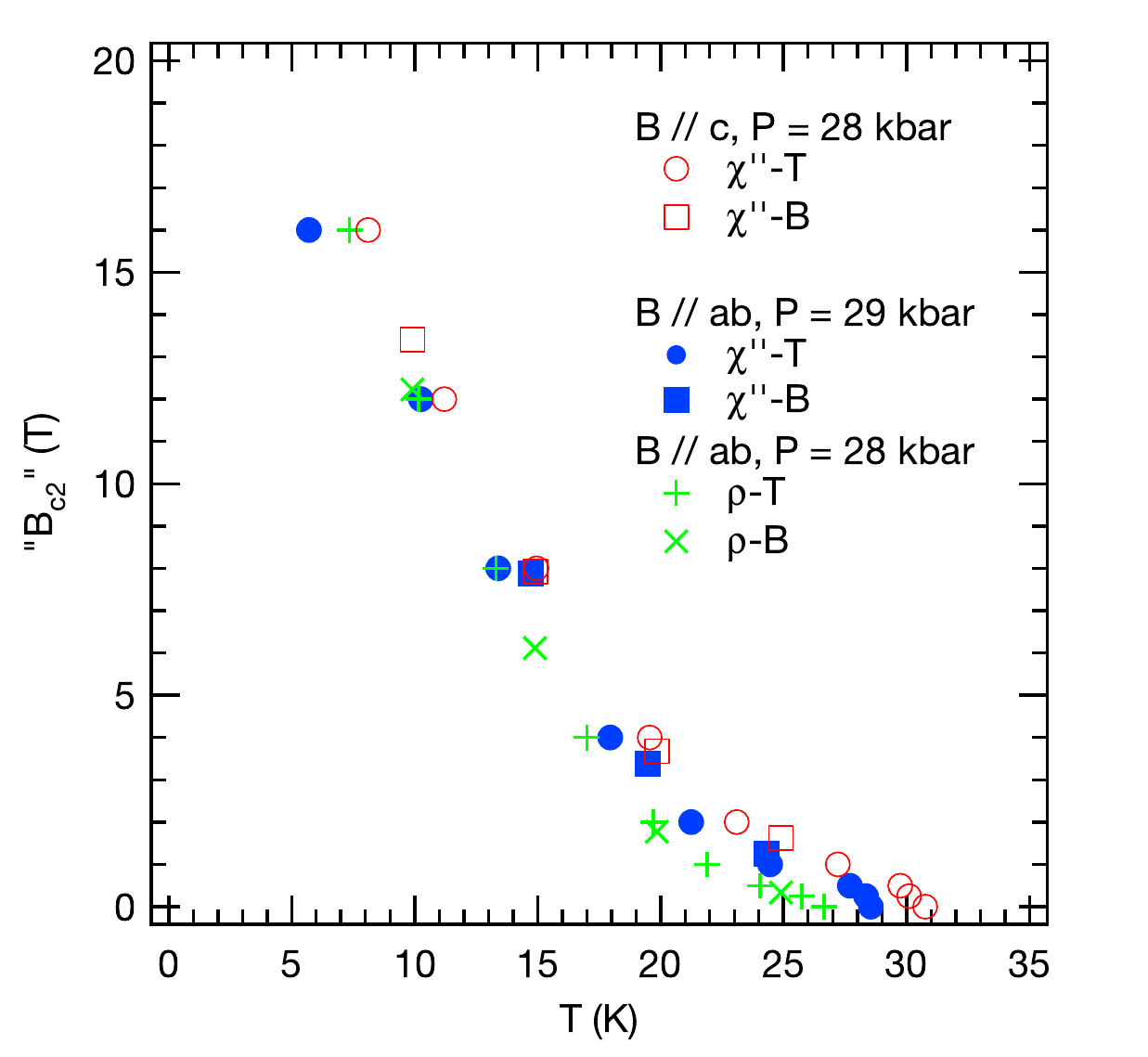}
\end{center}
\caption{(Color online) Upper critical field ``$B_{c2}$'' of EuFe$_2$As$_2$ under high $P$.  Note that the field values are based on the applied field $B_{appl}$ and are not corrected for the magnetization.}
\label{Bc2}
\end{figure}

Figure~\ref{chi_c}(a) shows $\chi_{ac}$ along the $c$-axis as a function of $T$ for various pressures and external fields.  At ambient $P$ ($P$ = 0 kbar) and $B_{appl}$ = 0 T, the real part $\chi^{\prime}$ increases with decreasing $T$ and reaches a maximum at $T_N$ = 20 K, below which the $T$ dependence is weak.  This is essentially the same as the $T$ dependence of dc magnetic susceptibility.\cite{Jiang09NJP}  At $P$ = 24 kbar and $B_{appl}$ = 0 T, $\chi^{\prime}$ exhibits a similar $T$ dependence with a maximum at 20 K, although it is enhanced over that at ambient $P$.  This is consistent with the previous report that $T_N$ is almost $P$-independent.\cite{Miclea08condmat}  We observed no clear indication of SC at $P$ = 24 kbar.  A small field of $B_{appl}$ = 0.25 T markedly changes the shape of the $\chi^{\prime}$-$T$ curve, and when $B_{appl}$ = 0.5 T no clear sign of $T_N$ is visible.  At $P$ = 28 kbar and $B_{appl}$ = 0 T, $\chi^{\prime}$ exhibits a large drop below $T_c$ = 31 K, then increases with decreasing $T$ to reach a maximum at 20 K, and finally decreases again.  The large drop in $\chi^{\prime}$ indicates the occurrence of SC, and the size of the drop is consistent with 100\% shielding, as indicated by the arrow in the figure.  The maximum at 20 K indicates that the AFM ordering of the Eu$^{2+}$ moments still occurs in the superconducting phase.  The feature associated with the magnetic transition is barely visible at $B_{appl}$ = 0.25 T, but is absent at $B_{appl}$ = 0.5 T.  With increasing $B_{appl}$, the superconducting transition shifts to lower $T$.

Figure~\ref{chi_c}(b) shows $\chi_{ac}$ along the $c$-axis as a function of $B_{appl}$ for various pressures and temperatures.  At $P$ = 0 kbar, $\chi^{\prime}$ shows a large drop at approximately 1.5 T, indicating the entrance into the field-induced paramagnetic state, as is consistent with a previous magnetization study.\cite{Jiang09NJP}  At $P$ = 24 kbar, $\chi^{\prime}$ at $B_{appl}$ = 0 is larger than that at $P$ = 0 kbar, as is consistent with the $\chi^{\prime}$-$T$ data explained above, and $\chi^{\prime}$ shows a drop at approximately 0.5 T.  The $\chi^{\prime}$-$B_{appl}$ curves at $P$ = 28 kbar show superconducting diamagnetism and its suppression with increasing $B_{appl}$.  The curves at $T$ = 0.02, 10, 15, and 20 K, namely, $T < T_N$, show a drop near 0.5 T, similar to the drop observed at $P$ = 24 kbar.

We define $T_c$ (``$B_{c2}$'') as the temperature (applied magnetic field) where $\chi^{\prime\prime}$ deviates from the normal-state value.  The reason why $\chi^{\prime\prime}$, not $\chi^{\prime}$, is used is that $\chi^{\prime}$ exhibits $T$ and field dependences in the normal state, which makes the unambiguous determination of the onset of SC difficult in some cases [see the $B_{appl}$ = 1 T curve in Fig.~\ref{chi_c}(a) and the $T$ = 25 K curve in Fig.~\ref{chi_c}(b), for example].  The quotation marks attached to $B_{c2}$ indicate that $B_{c2}$ values are based on the applied field, not on the internal field.  The superconducting phase diagram determined from the data in Fig.~\ref{chi_c} is shown in Fig.~\ref{Bc2}.

Figure~\ref{chi_ab}(a) shows $\chi_{ac}$ in the $ab$-plane at $P$ = 29 kbar as a function of $T$.  With decreasing $T$, the $B_{appl}$ = 0 T curve shows a large drop at $T_c$ = 29 K, then increases to reach a plateau between $\sim$20 and 15 K, and finally decreases.  The magnitude of diamagnetism is nearly 100\%.  Although this is not typical of antiferromagnets, the plateau indicates a magnetic phase transition.  Note that the $T$ dependence of the magnetic susceptibility at ambient $P$ shows a similar plateau when measured at $B_{appl}$ = 0.5 T.\cite{Jiang09NJP}  An indication of a magnetic phase transition is barely visible at $B_{appl}$ = 0.25 T, but is absent at 0.5 T.  Figure~\ref{chi_ab}(b) shows $\chi_{ac}$ in the $ab$-plane at $P$ = 29 kbar as a function of $B_{appl}$.  The curves at $T$ = 0.3, 5, 10, 15, and 20 K, namely $T < T_N$, show a decrease with increasing field up to $\sim$1 T, similarly to the $c$-axis data.  The superconducting phase diagram is shown in Fig.~\ref{Bc2}.

Figure~\ref{rho_curves}(a) shows $\rho$ at $P$ = 25 and 28 kbar as a function of $T$.  At $P$ = 25 kbar, a hump appears at approximately 100 K, indicating the existence of $T_o$.  $\rho$ shows a sudden drop below 30 K and a hump at approximately 20 K, but does not reach zero (see the inset).  This is similar to the data in ref.~\citen{Miclea08condmat}.  At $P$ = 28 kbar, $\rho$ decreases almost linearly with decreasing $T$ from RT and exhibits a sharp drop to zero below 30 K.  The midpoint $T_c$ is 29 K.  Figures~\ref{rho_curves}(b) and \ref{rho_curves}(c) show resistive transition curves.  The transition width increases as $B_{appl}$ increases.  To determine the superconducting phase diagram (Fig.~\ref{Bc2}), we define $T_c$ (``$B_{c2}$'') as the temperature (applied magnetic field) where $\rho$ deviates from zero.  This criterion gives a much better correspondence with magnetically determined values of $T_c$ at large values of $B_{appl}$, where the resistive transition widths are large, than the usual midpoint one, though it gives a lower $T_c$ of 27 K at $B_{appl}$ = 0 T.

Although ref.~\citen{Miclea08condmat} suggested that $P$-induced SC coexists with the structural/magnetic transition at $T_o$, the present $\chi_{ac}$ and $\rho$ data indicate that \textit{bulk} SC occurs only when $T_o$ is completely suppressed.  There is evidence from the $\mu$SR and NMR measurements of (Ba$_{1-x}$K$_x$)Fe$_2$As$_2$ that the apparent coexistence of the orthorhombic AFM phase and superconducting phase in iron pnictide superconductors is not a true coexistence but can be explained by phase separation.\cite{Park09PRL, Fukazawa09JPSJ}  Also note that the critical pressure at which the structural/magnetic transition disappears differs between ref.~\citen{Miclea08condmat} and the present work:  in ref.~\citen{Miclea08condmat}, no transition was detected above $P$ = 23 kbar, while a transition was detected at $P$ = 25 kbar in our study.  A very recent high-$P$ study of SrFe$_2$As$_2$ showed that the critical pressure is very sensitive to the homogeneity of the applied pressure and that it is higher when the pressure is more hydrostatic.\cite{Kotegawa09condmat}  This may explain the difference in the critical pressure between ref.~\citen{Miclea08condmat} and the present work.  As shown in Fig.~\ref{rho_curves}(a), the appearance of bulk SC is associated with the anomalous nearly $T$-linear $\rho$.  This and the fact that partial (or filamentary) and bulk SC occur below and above the critical pressure of magnetism, respectively, bear resemblance to cases of $P$-induced SC in some heavy-fermion compounds.\cite{Kimura05PRL, Knebel08JPSJ}  It is interesting to note that $T$-linear $\rho$ is also observed in a wide $T$ range in optimally doped BaFe$_{1.8}$Co$_{0.2}$As$_2$.\cite{Ahilan08JPCM}

The superconducting phase diagram (Fig.~\ref{Bc2}) indicates that ``$B_{c2}$'' is almost isotropic.  The initial slope -d``$B_{c2}$''/d$T$ at $T$ = $T_c$ can be estimated to be 0.5(1), 0.6(2), and 0.19(4) T/K from the $c$-axis $\chi_{ac}$, $ab$-plane $\chi_{ac}$, and $ab$-plane $\rho$ data for $B_{appl}\leqslant0.5$ T, respectively.  Although the slope depends on the type of determination method for ``$B_{c2}$'', the estimated slopes are much smaller than those in the $P$-induced SC of other 122 compounds.\cite{Torikachvili08PRL, Park08JPCM, Kotegawa09JPSJ}  It is interesting to note that a very large slope of 3.87 T/K was reported for Eu$_{0.7}$Na$_{0.3}$Fe$_2$As$_2$.\cite{Qi08NJP}  The ``$B_{c2}$''-$T$ curves show a strong concave curvature above about 1 T.  This reminds us of the $T$ dependence of $B_{c2}$ observed in the ternary molybdenum sulphide Sn$_{0.2}$Eu$_{0.8}$Mo$_{6.35}$S$_8$,\cite{FischerJPCSSC75} which can be explained by the Jaccarino-Peter compensation effect arising from the exchange interaction between local moments and conduction carriers.\cite{Jaccarino62PRL}
 
In conclusion, when the structural/magnetic transition is suppressed by high $P$, EuFe$_2$As$_2$ shows an anomalous nearly $T$-linear dependence of $\rho$ and becomes a bulk superconductor at $T_c \sim$30 K.  The AFM order of the Eu$^{2+}$ moments at $T_N \sim$20 K persists in the superconducting phase.  The upper critical field exhibits a unique $T$ dependence, which indicates the effect of the exchange interaction between the Eu$^{2+}$ moments and conduction carriers.


\begin{thebibliography}{10}
\bibitem{Daphne7474}
Daphne7474 was developed by Idemitsu Kosan Co. Ltd. in collaboration with one of the
  authors KM for use in high-pressure~scientific researches.
\end{thebibliography}
\end{document}